\documentclass[aps,prb,floatfix,twocolumn,showpacs]{revtex4}%
\usepackage{amsmath}
\usepackage{graphicx}
\usepackage{amsfonts}
\usepackage{amssymb}%
\providecommand{\U}[1]{\protect\rule{.1in}{.1in}}
\providecommand{\U}[1]{\protect\rule{.1in}{.1in}}
\providecommand{\U}[1]{\protect\rule{.1in}{.1in}}
\providecommand{\U}[1]{\protect\rule{.1in}{.1in}}

\begin{document}
\title{Cavity photons as a probe for charge relaxation resistance and photon emission
in a quantum dot coupled to normal and superconducting continua}
\author{L.E. Bruhat$^{1}$, J.J. Viennot$^{1,2}$, M.C. Dartiailh$^{1}$, M.M.
Desjardins$^{1}$, T. Kontos$^{1}$ and A. Cottet$^{1}$}
\affiliation{$^{1}$Laboratoire Pierre Aigrain, Ecole Normale Sup\'{e}rieure-PSL Research
University, CNRS, Universit\'{e} Pierre et Marie Curie-Sorbonne
Universit\'{e}s, Universit\'{e} Paris Diderot-Sorbonne Paris Cit\'{e}, 24 rue
Lhomond, F-75231 Paris Cedex 05, France}
\affiliation{$^{2}$JILA and the Department of Physics, University of Colorado, Boulder,
Colorado 80309, USA}

\pacs{42.50.Pq, 74.25.N-,73.23.-b, 73.63.Fg}
\date{\today}

\begin{abstract}
Microwave cavities have been widely used to investigate the behavior of closed
few-level systems. Here, we show that they also represent a powerful probe for
the dynamics of charge transfer between a discrete electronic level and
fermionic continua. We have combined experiment and theory for a carbon
nanotube quantum dot coupled to normal metal and superconducting contacts. In
equilibrium conditions, where our device behaves as an effective quantum
dot-normal metal junction, we approach a universal photon dissipation regime
governed by a quantum charge relaxation effect. We observe how photon
dissipation is modified when the dot admittance turns from capacitive to
inductive. When the fermionic reservoirs are voltage biased, the dot can even
cause photon emission due to inelastic tunneling to/from a
Bardeen-Cooper-Schrieffer peak in the density of states of the superconducting
contact. We can model these numerous effects quantitatively in terms of the
charge susceptibility of the quantum dot circuit. This validates an approach
that could be used to study a wide class of mesoscopic QED devices.

\end{abstract}
\maketitle

\section{Introduction}

Circuit QED techniques\cite{Wallraff} have been recently put forward to
investigate the electronic dynamics in quantum dot
circuits\cite{Childress,Delbecq1,Frey1}, or, more generally, mesoscopic
circuits\cite{Cottet2015}. So far, the interaction between cavity photons and
charges\cite{Petersson,Schroer,Frey2,Toida1,Basset,Zhang,Viennot,Liu,Liu2,Stockklauser,Liu3}
or spins\cite{Viennot2} confined in quantum dots has raised most experimental
attention. This atomic-like limit is a priori very appealing for quantum
information applications since it goes, in principle, with long coherence
times. Nevertheless, mesoscopic circuits are inseparable from the existence of
electronic reservoirs with Fermi seas. These fermionic reservoirs are not
necessarily a drawback. For instance, a strong coupling between a dot and a
normal metal enables to revisit condensed matter problems such as the Kondo
effect. Ferromagnetic contacts can be used to design spin quantum
bits\cite{Viennot2}, or study spin-dependent transport\cite{Cottet}.
Superconducting contacts are crucial for the study of Cooper pair
splitting\cite{Hofstetter,Herrmann}, Andreev bound
states\cite{Pillet,Dirks2,Janvier}, and Majorana quasiparticles\cite{Mourik}.
In principle, microwave cavities could represent a powerful tool to
investigate these
features\cite{Skoldberg,CottetCPS1,CottetCPS2,Schmidt,Trif,CottetMajos,Dmytruk0,Chirla}%
.

In this context, it is crucial to understand how tunneling processes between a
discrete energy level and the continuum of states of a reservoir can affect
cavity photons. This situation is epitomized by a single quantum dot circuit
coupled to a cavity, a case which has been studied elusively so
far\cite{Delbecq1,Delbecq2,Frey2}. A recent experiment has revealed that the
quantum dot can add an effective capacitance or an inductance to the photons
environment, depending on the transparency of its contacts\cite{Frey2}.
However, the cavity dissipation expected together with this effect has been
left unexplored. On the theory side, most experiments combining quantum dot
circuits and microwave resonators have been interpreted by disregarding
fermionic reservoirs or by using a Lindbladt equation suitable for
dot/reservoir tunnel rates much smaller than the temperature of the
experiment. An alternative approach is highly desirable for investigating the
open contacts limit. Descriptions in terms of the charge susceptibility of the
quantum dot circuits have been recently
suggested\cite{Cottet2015,Cottet:11,Schiro,Dmytruk}.

In this work, we study experimentally and theoretically the behavior of a
single quantum dot in a carbon nanotube, coupled to normal metal (N) and
superconducting (S) reservoirs, and embedded in a high finesse microwave
cavity. In a first step, we study a dot with a discrete level coupled only to
the N reservoir. In this case, the current response of the dot to a direct
gate voltage excitation $V_{RF}$ can be developed as $I_{RF}=i\omega_{RF}%
C_{Q}(1-i\omega_{RF}R_{AC}C_{Q})V_{RF}+o(\omega_{RF}^{2})$, provided the
frequency $\omega_{RF}$ of the excitation is smaller than the tunnel rate
$\Gamma_{N}$ of the N/dot junction. This development maps onto that expected
for a RC circuit with capacitance $C_{Q}$ and resistance $R_{AC}$. The
capacitance $C_{Q}$ characterizes the ability of the dot to host electrons at
DC or low frequencies. The resistance $R_{AC}$ describes the dynamics of
charge tunneling through the junction, which leads to the relaxation of the
charge imbalance caused by $V_{RF}$. For an incoherent device, $R_{AC}$
depends on the detailed properties of the dot circuit. However, in the
coherent non-interacting limit with $\omega_{RF}\ll\Gamma_{N}$, B\"{u}ttiker
and coworkers have predicted that $R_{AC}$ should take a universal value,
independently of the dot parameters\cite{Buttiker,Pretre}. This effect, which
reveals the wavy behavior of electrons inside the dot, is of fundamental as
well as practical importance in the context of the miniaturization of
electronic devices. The only observation of a universal $R_{AC}$ so far was
performed by Gabelli et, al. with a strongly spin-polarized GaAs quantum dot,
using AC conductance measurements\cite{Gabelli}. Remarkably, this dot was in a
non-interacting regime due to a top gate with an unusually large
capacitance\cite{Feve}. Noticeably, the independence of $R_{AC}$ from the dot
orbital energy was not tested by Gabelli and coworkers. In our system, this
property appears as a scaling between the dissipation and dispersion induced
by the dot on the cavity. We can confirm experimentally this scaling behavior
for intermediate tunnel rates $\Gamma_{N}\sim2.5\omega_{RF}$ where it is
already approximately valid and well resolvable. Remarkably, this effect
occurs in spite of the presence of strong Coulomb blockade in our sample. The
effect of interactions on quantum charge relaxation has raised an intense
theoretical activity because, in practice, most quantum dots are subject to
strong Coulomb interactions. In this limit, predictions for $R_{AC}$ display a
rich phenomenology\cite{RC1,RC3,RC4,RC5,Minchul,Rodionov,Burmistrov}.
Nevertheless, it was recently suggested that a universal charge relaxation
resistance persists in the spin-degenerate interacting case\cite{Minchul,RC5}.
Our results are consistent with this prediction. Finally, we observe how
quantum charge relaxation depends on the dot orbital energy for smaller tunnel
rates, when the dot admittance turns from capacitive to inductive.

In a second step, we study the finite bias voltage regime where the dot level
is also coupled to the S reservoir. Contextually, the implementation of
mesoscopic QED experiments with superconducting hybrid circuits is very
recent. Atomic contacts between superconductors have been used to form a new
type of quantum bit based on Andreev bound states\cite{Janvier}.
Semiconducting nanowires have been used for realizing Josephson junctions in
superconducting circuits\cite{Larsen,Lange}. However, quantum dot circuits
with superconducting reservoirs have been coupled neither to microwave
cavities nor to a direct AC\ excitation, so far. Despite this lack of
experiments, photon-assisted tunneling between a dot and a superconductor has
raised theoretical interest for more than 15
years\cite{Zhao,Sun2,Nurbawono,Whan,Cho,Nguyen,
Zhu,Moghaddam,Avriller,Baranski}. The coupling between superconductor/quantum
dot hybrid circuits and microwave cavities has also been studied in recent
theory works\cite{Skoldberg,CottetCPS1,CottetCPS2}. Here, we show
experimentally that a microwave cavity is able to reveal photon-assisted
tunnel events between a dot and the BCS peaks of a superconductor, not visible
in the dot current. In particular, we observe negative photon damping, which
reveals photon emission. This result illustrates that quasiparticle tunneling
to a fermionic reservoir does not always induce photonic dissipation.

In order to understand our measurements, we use a Keldysh Green's function
approach. We can reproduce simultaneously the quantum dot conductance and the
microwave response of the cavity, versus the dot gate and bias voltages, with
an unprecedented accuracy for this type of hybrid system. We thereby validate
the description of mesoscopic QED experiments in terms of an electronic charge
susceptibility. To illustrate the broad scope of this approach, we present its
multi-dot generalization, which can be used for many different geometries,
like for instance Cooper pair splitters and topological hybrid nanocircuits.
Our work also opens wide experimental perspectives since microwave cavities
appear as a powerful probe for quantum charge relaxation, photon-assisted
tunneling, and all other effects involving tunneling between a discrete level
and fermionic reservoirs.

This article is organized as follows. Section \ref{ES} presents our
experimental setup. Sections \ref{LB} and \ref{HB} analyze our experimental
data, for the N/dot and N/dot/S limits respectively. Section \ref{di}
summarizes our results, presents the multi-dot generalization of our approach,
and various perspectives. Appendix A shows experimental details and
supplementary data. Appendix B discusses our theoretical approach.

\begin{figure}[t]
\includegraphics[width=1.\linewidth]{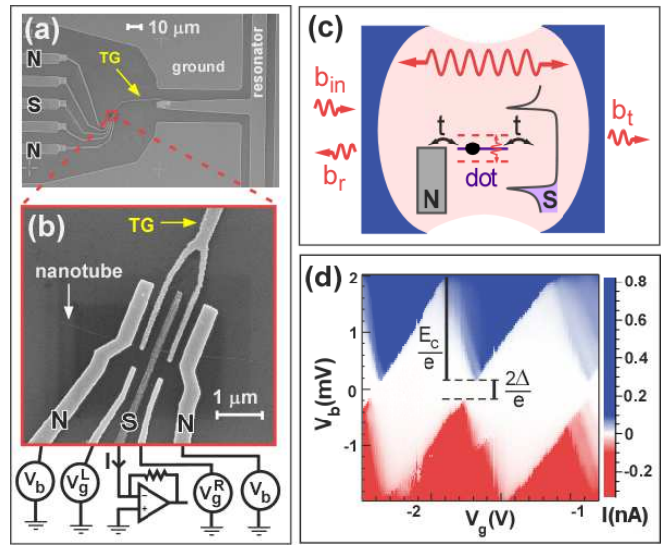}\caption{Panels (a) and (b):
Scanning electron micrograph of the microwave resonator and the quantum dot
circuit. Panel (c): Principle of our setup. The dot level is tunnel coupled to
the N and S reservoirs and modulated by the cavity electric field. Panel (d):
Current through the S contact versus the effective gate voltage $V_{g}$ and
the bias voltage $V_{b}$.}%
\label{setup}%
\end{figure}

\section{Experimental setup\label{ES}}

We use a carbon nanotube on which we evaporate a superconducting contact (S)
surrounded by two normal metal (N) contacts, visible in Fig.1b (technical
details are presented in Appendix A). In our regime of parameters, the whole
nanotube section between the two N contacts forms a single quantum dot. An
effective gate voltage $V_{g}$ is used to tune the dot level orbital energy
$\varepsilon_{d}$. We connect the S contact to ground and we apply the same
bias voltage $V_{b}$ to the two N contacts, which can thus be considered as an
effective single contact\cite{sc}. The dot is capacitively coupled to the
central conductor of a superconducting coplanar waveguide cavity through a top
gate TG (see Fig.\ref{setup}a). We measure the cavity transmission
$b_{t}/b_{in}$ at a frequency $\omega_{RF}$ equal to the bare cavity frequency
$\omega_{0}\sim2\pi\times6.65$~\textrm{GHz}. We determine the phase shift
$\Delta\varphi$ and the reduced amplitude shift $\Delta A/A_{0}$ of
$b_{t}/b_{in}$, which are caused by the presence of the quantum dot circuit,
with $A_{0}$ the bare cavity transmission amplitude. Simultaneously, we
measure the DC current $I$ and differential conductance $G$ through the dot.
The current $I$ shows clear signatures of Coulomb blockade with a charging
energy $E_{c}\approx1.8~\mathrm{meV}$ (see Fig.\ref{setup}d). It also vanishes
for a bias voltage $V_{b}$ smaller than the gap $\Delta\simeq0.17~\mathrm{meV}%
$ of the S contact. Therefore, for $V_{b}=0$, the effect of the S contact can
be disregarded and the quantum dot circuit corresponds to an effective N/dot
junction, studied in section \ref{LB}. For $e\left\vert V_{b}\right\vert
>\Delta$, our device enables the study of quasiparticle transport in a N/dot/S
bi-junction, presented in section \ref{HB}.

\section{Photon dissipation in an effective N/dot junction\label{LB}}

\begin{figure}[t]
\includegraphics[width=1.\linewidth]{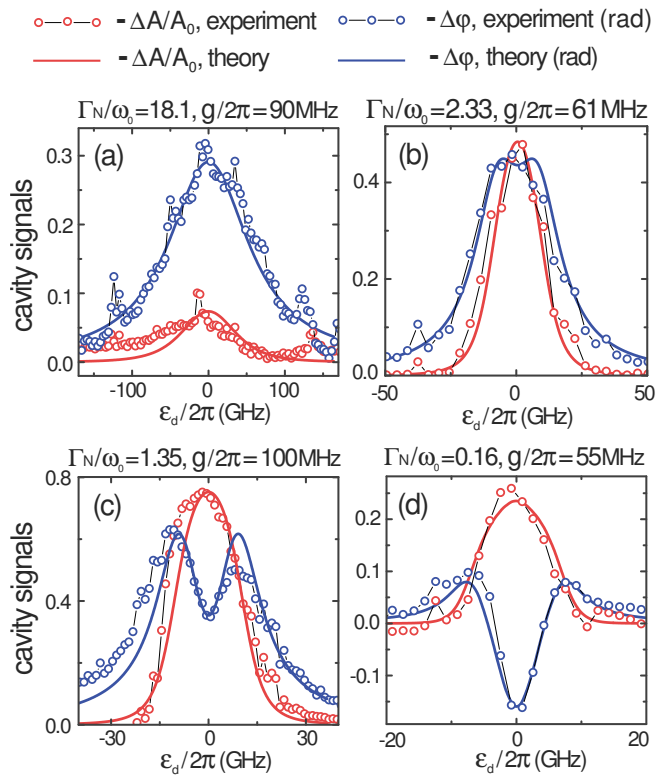}\caption{Measured phase
shift $\Delta\varphi$ (blue dots) and reduced amplitude shift $\Delta A/A_{0}$
(red dots) of the microwave signal transmitted by the cavity versus the energy
$\varepsilon_{d}$ of the dot orbital, for $V_{b}=0$ and different dot orbitals
with decreasing tunnel rates $\Gamma_{N}$ \ from top to bottom and left to
right panels (for clarity, we have plotted the opposite of these signals). The
red and blue lines show the predictions given by Eqs. (\ref{XiT}) and
(\ref{rr}) for values of $\Gamma_{N}$ and $g$\ given in the different panels
and $T=60$ $\mathrm{mK}\simeq0.19\omega_{0}$.}%
\label{MosaikLittle.eps}%
\end{figure}

A single dot level coupled to a N reservoir is the most basic configuration
for studying the light matter interaction in a mesoscopic circuit. Our device
realizes such a situation for $V_{b}=0$ due to the absence of subgap Andreev
reflections. Figure \ref{MosaikLittle.eps} shows the cavity signals
$\Delta\varphi$ (blue dots) and $\Delta A/A_{0}$ (red dots) versus the energy
$\varepsilon_{d}$ of the dot orbital, for $V_{b}=0$, and decreasing tunnel
rates $\Gamma_{N}$ from left to right and top to bottom panels. The
correspondence between $\varepsilon_{d}$ and the gate voltage $V_{g}$ is given
in appendix A for each dot level. We observe resonances although $I=0$. This
means that the cavity is able to reveal quasiparticle tunneling between the
dot and the N contact even if it does not lead to a DC current. The phase
signal $\Delta\varphi$ (blue dots) can be positive as well as negative, as
already observed in Ref.\cite{Frey2}, depending on the value of $\Gamma_{N}$.
This is because, for $\omega_{0}\ll\Gamma_{N}$, the quantum dot circuit
behaves as an effective capacitance. Electrons can follow very rapidly the
variations of the dot potential to go in and out of the dot, proportionally to
the dot density of states. However, for $\omega_{0}\gg\Gamma_{N}$, the charge
current lags behind the dot potential, so that the dot behavior becomes
inductive\cite{Buttiker,Wang}. In contrast, the signal $\Delta A$ (red dots)
always remains negative, up to experimental uncertainty. One could naively
expect that $\Delta A$, which reveals photon dissipation, will scale with
$\Gamma_{N}$ which is the main dissipation parameter in our problem. However,
this intuition is wrong since $\Delta A$ becomes small when $\Gamma_{N}$ tends
to large values (see Fig.\ref{MosaikLittle.eps}a).

To understand the behavior of our device, we use the Hamiltonian%
\begin{align}
H_{tot}  &  =H_{d}+\omega_{0}\hat{a}^{\dag}\hat{a}+g(\hat{a}+\hat{a}^{\dag
})\sum\limits_{\sigma}\hat{d}_{\sigma}^{\dag}\hat{d}_{\sigma}\nonumber\\
&  +\sum\limits_{p}\omega_{p}\hat{b}_{p}^{\dag}\hat{b}_{p}+\sum\limits_{p}%
(\tau_{p}\hat{b}_{p}^{\dag}\hat{a}+\tau_{p}^{\ast}\hat{a}^{\dag}\hat{b}_{p})
\label{Htot}%
\end{align}
where $H_{d}$ describes the quantum dot circuit (see appendix B for details),
$\hat{d}_{\sigma}^{\dag}$ adds an electron with spin $\sigma$ in the dot
level, $\hat{a}^{\dag}$ adds a photon in the cavity and $\hat{b}_{p}^{\dag}$
describes a bosonic bath which accounts for the cavity intrinsic linewidth
$\Lambda_{0}$. We assume that cavity photons modulate the chemical potential
of the quantum dot with a coupling constant $g=e\varkappa V_{rms}$, with
$V_{rms}$ the cavity root mean square voltage and $e$ the electron charge. The
dimensionless coupling constant $\varkappa$ depends on the overlap between the
electron wavefunction associated with the dot level and the photonic
pseudopotential, which is spatially non-uniform \cite{Cottet2015}. Therefore,
the value of $g$ generally depends on the dot level considered, as we will see
in the experimental data. Using Eq.(\ref{Htot}), a semiclassical linear
response approach leads to the cavity transmission (see Appendix B):%
\begin{equation}
\frac{b_{t}}{b_{in}}=\frac{t_{0}}{\omega_{RF}-\omega_{0}-i\Lambda_{0}%
-g^{2}\chi(\omega_{0})} \label{InOut}%
\end{equation}
The quantum dot charge susceptibility $\chi(\omega)$ can be calculated within
the Keldysh formalism as%
\begin{equation}
\chi^{\ast}(\omega)=-i%
%TCIMACRO{\tint }%
%BeginExpansion
{\textstyle\int}
%EndExpansion
\frac{d\omega}{2\pi}\mathrm{Tr}\left[  \mathcal{\check{S}}(\omega
)\mathcal{\check{G}}^{r}(\omega)\check{\Sigma}^{<}(\omega)\mathcal{\check{G}%
}^{a}(\omega)\right]  \label{Xitot}%
\end{equation}
with%
\begin{equation}
\mathcal{\check{S}}(\omega)=\check{\tau}\left(  \mathcal{\check{G}}^{r}%
(\omega+\omega_{0})+\mathcal{\check{G}}^{a}(\omega-\omega_{0})\right)
\check{\tau} \label{SSS}%
\end{equation}
The retarded and advanced Green's functions $\mathcal{\check{G}}^{r/a}$ of the
quantum dot and the lesser self energy $\check{\Sigma}^{<}(\omega)$ are
defined in Appendix B. The matrix $\check{\tau}=diag(1,-1)$ describes the
structure of the photon/particle coupling in the Nambu (electron/hole) space.
Note that this degree of freedom is not necessary for describing the N/dot
junction, but we introduce it for a later use in section \ref{HB}. In the
present section, we disregard the S reservoir and use $E_{c}=0$ so that the
susceptibility $\chi(\omega)$ can be simplified as Eq. (\ref{Xi0}) of Appendix
B at zero temperatures and%
\begin{equation}
\chi(\omega)=%
%TCIMACRO{\dint _{-\infty}^{+\infty}}%
%BeginExpansion
{\displaystyle\int_{-\infty}^{+\infty}}
%EndExpansion
\frac{d\omega^{\prime}}{\pi\omega}\frac{\Gamma_{N}\left(  f(\omega^{\prime
})-f(\omega^{\prime}-\omega)\right)  }{(\omega^{\prime}-\varepsilon_{d}%
-i\frac{\Gamma_{N}}{2})(\omega^{\prime}-\omega-\varepsilon_{d}+i\frac
{\Gamma_{N}}{2})} \label{XiT}%
\end{equation}
with $f(\varepsilon)=1/(1+\exp[\varepsilon/k_{b}T])$ for finite temperatures.
Below, we interpret our data by using the exact non-interacting expression
(\ref{XiT}) of $\chi(\omega)$, which depends only on two parameters: the
tunnel rate $\Gamma_{N}$ between the dot and N and the temperature $T$. We
obtain a quantitative agreement between the measured $(\Delta\varphi,\Delta
A)$ and the values calculated from the transmission ratio
\begin{equation}
(1+(\Delta A/A_{0}))e^{i\Delta\varphi}=\Lambda_{0}/(\Lambda_{0}-ig^{2}%
\chi(\omega_{0})) \label{rr}%
\end{equation}
which follows from Eq.(\ref{InOut}) for $\omega_{RF}=\omega_{0}$ (see red and
blue lines in Fig. \ref{MosaikLittle.eps}). We use the same finite temperature
$T=60~\mathrm{mK}$ for all the resonances. Then, for each resonance, there
remains only two adjustable parameters, namely $g$ and $\Gamma_{N}$, to fit
simultaneously the $\Delta\varphi$ and $\Delta A$ curves. Remarkably, we
obtain a good agreement with the data for a wide range of $\Gamma_{N}%
/\omega_{0}$ ratios (see Fig. \ref{MosaikBig} for supplementary resonances).
The full functional form of the cavity response is accurately reproduced by
our theory. Such a modeling was not possible for previous experiments
combining (real or effective) single quantum dots with microwave
cavities\cite{Delbecq1,Delbecq2,Frey2}.\begin{figure}[t]
\includegraphics[width=1.\linewidth]{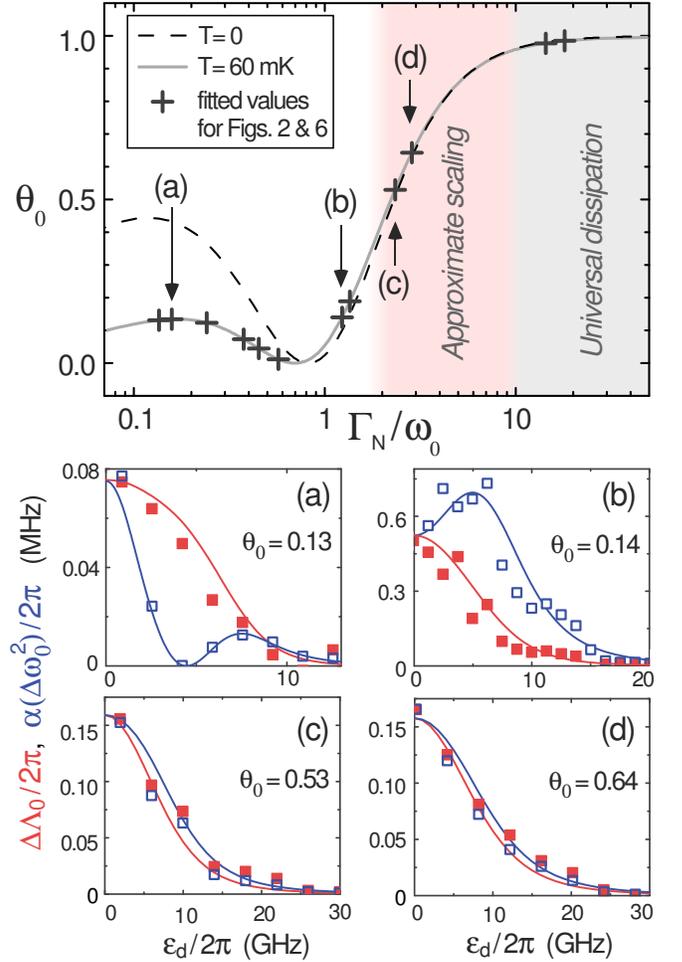}\caption{{}Top panel: Ratio
$\theta_{0}=\theta(\varepsilon_{d}=0)$ versus the tunnel rate $\Gamma_{N}$,
calculated from Eqs. (\ref{XiT}), (\ref{theta}), and (\ref{Xi0}) for $T=0$
(black dashed line) and $T=60~$\textrm{mK} (gray full line). The crosses
correspond to fitted values of $\theta_{0}$, calculated from Eqs. (\ref{XiT})
and (\ref{theta}) for the different resonances in Figs. \ref{MosaikLittle.eps}
and \ref{MosaikBig}. Bottom panels: Comparison between the experimental
$\Delta\Lambda_{0}$ and ($\Delta\omega_{0}$)$^{2}$, using the scaling factor
$\alpha=\pi\omega_{0}/2\theta_{0}g^{2}$ , with $\theta_{0}$ indicated with
arrows in the top panel. We use $\Gamma_{N}/\omega_{0}=0.16$, $1.23$, $2.33$
and $2.86$ from left to right and top to bottom panels. We also show as blue
and red full lines the calculated $\Delta\Lambda_{0}$ and ($\Delta\omega_{0}%
$)$^{2}$ .}%
\label{ratio}%
\end{figure}

A deeper analysis of the cavity response can be performed by studying the
cavity frequency shift $\Delta\omega_{0}$ and cavity linewidth shift
$\Delta\Lambda_{0}$, which can be obtained from the experimental signals
as\cite{mapping,noteS} $\Delta\omega_{0}=\Lambda_{0}(A_{0}/A)\sin
(\Delta\varphi)$ and $\Delta\Lambda_{0}=\Lambda_{0}\left(  (A_{0}%
/A)\cos(\Delta\varphi)-1\right)  $ and modeled theoretically from
$\Delta\omega_{0}+i\Delta\Lambda_{0}=g^{2}\chi(\omega_{0})$. To study the
relation between $\Delta\Lambda_{0}$ and $\Delta\omega_{0}$, we define the
ratio%
\begin{equation}
\theta=\frac{\pi}{2}\frac{\omega_{0}}{g^{2}}\frac{(\Delta\omega_{0})^{2}%
}{\Delta\Lambda_{0}} \label{thetaexp}%
\end{equation}
which can be modeled theoretically as%
\begin{equation}
\theta=\frac{\pi}{2}\omega_{0}\frac{(\operatorname{Re}[\chi])^{2}%
}{\operatorname{Im}[\chi]} \label{theta}%
\end{equation}
The top panel of Fig.\ref{ratio} shows with a dashed line $\theta_{0}%
=\theta(\varepsilon_{d}=0)$ versus $\Gamma_{N}/\omega_{0}$, calculated at
$T=0$ from Eqs.(\ref{theta}) and (\ref{Xi0}), for a dot level at resonance
with the Fermi energy of the reservoir ($\varepsilon_{d}=0$). Remarkably,
$\theta_{0}$ shows the minimum $\theta_{0}=0$ for $\Gamma_{N}\sim0.7\omega
_{0}$ due to the inductive to capacitive crossover of
Fig.\ref{MosaikLittle.eps}. Then, in the adiabatic limit $\Gamma_{N}\gg
\omega_{0}$, $\theta_{0}$ tends to 1. In fact, this limit is valid for any
value of $\varepsilon_{d}$, i.e.
\begin{equation}
\lim\limits_{\Gamma_{N}/\omega_{0}\rightarrow+\infty}\theta(\varepsilon_{d})=1
\label{limit}%
\end{equation}
The full gray line in Fig.\ref{ratio}, top panel, shows $\theta_{0}$ for the
temperature $T=60~\mathrm{mK}$, calculated from Eqs. (\ref{XiT}) and
(\ref{theta}). It illustrates that finite temperatures affect quantitatively
the behavior of the system for low values of $\Gamma_{N}$, but Eq.(\ref{limit}%
) remains valid as soon as $\Gamma_{N}\gg k_{B}T$. A straightforward question
is whether the non-trivial behavior of Eq.(\ref{limit}) can be observed with
our experiment. This equation has two important implications. First, it
predicts that the $\Delta\Lambda_{0}$ and $(\Delta\omega_{0})^{2}$ curves
versus $\varepsilon_{d}$ (or equivalently versus the dot gate voltage $V_{g}$)
should be proportional in the open contact limit. Second, it gives the exact
value of the proportionality constant between $\Delta\Lambda_{0}$ and
$(\Delta\omega_{0})^{2}$. The latter cannot be accessed in our experiment.
Indeed, we cannot calibrate the absolute value of $\theta$ because we don't
have an independent experimental determination of the parameter $g$. Instead,
we determine $g$ and thus $\theta$ from a fitting procedure which relies on
the assumptions of our theory. Nevertheless, we can test experimentally the
scaling between $\Delta\Lambda_{0}$ and $(\Delta\omega_{0})^{2}$, as discussed below.

To illustrate the large variety of regimes achieved with our experiment, we
show with crosses the fitted values of $\theta_{0}$, calculated from Eqs.
(\ref{XiT}) and (\ref{theta}), for the fitting parameter $\Gamma_{N}$ of the
different resonances in Figs.\ref{MosaikLittle.eps} and \ref{MosaikBig} and
$T=60~\mathrm{mK}$. In principle, the scaling between $\Delta\Lambda_{0}$ and
$(\Delta\omega_{0})^{2}$ should be closely satisfied in the gray area where
$\theta_{0}\simeq1$. However, for the two resonances we have found in this
area ($\Gamma_{N}=18.1\omega_{0}$ and $\Gamma_{N}=14.3\omega_{0}$), we cannot
determine reliably $\Delta\Lambda_{0}$ from $\Delta A$ and $\Delta\varphi$
because $\Delta A$ is small and thus too much affected by background
variations. This difficulty raises because in the adiabatic limit, the dot
charge is in phase with the dot gate excitation, i.e. $\chi(\omega_{0}%
\ll\Gamma_{N})=\hbar\partial\left\langle n\right\rangle /\partial
\varepsilon_{d}\in\mathbb{R}$ with $\left\langle n\right\rangle $ the static
charge occupation of the dot. This is why, for $\Gamma_{N}\gg\omega_{0}$, we
find that $\Delta\Lambda_{0}=g^{2}\operatorname{Im}[\chi(\omega_{0}%
)]\sim8g^{2}(\omega_{0}/\Gamma_{N})^{2}/\pi\omega_{0}$ vanishes like
$(\omega_{0}/\Gamma_{N})^{2}$. From Ref.\cite{noteS}, $\Delta A$ is itself
small in this case. Nevertheless, we can interpret the raw cavity signals
$\Delta\varphi$ and $\Delta A$ with the same theory as our other data, which
shows that they are consistent with the universality of charge relaxation (see
Figs.\ref{MosaikLittle.eps}a and \ref{MosaikBig}). To resolve the scaling
behavior of the cavity response, we now consider the resonances at $\Gamma
_{N}=2.40\omega_{0}$ and $\Gamma_{N}=2.86\omega_{0}$. These points belong to
the pink area $2\leq\Gamma_{N}\lesssim10$ of Fig.\ref{ratio}, where, from our
theory, the scaling behavior should still hold approximately, although
$\theta_{0}<1$. As visible in panels \ref{ratio}c and \ref{ratio}d, we indeed
find that the proportionality between the experimental $\Delta\Lambda_{0}$ and
$(\Delta\omega_{0})^{2}$ is satisfied to a good accuracy, with a scaling
factor $\alpha=\pi\omega_{0}/2\theta_{0}g^{2}$. Small discrepancies between
$\Delta\Lambda_{0}$ and $\alpha(\Delta\omega_{0})^{2}$, are visible in the
theoretical curves (see red and blue lines) but not resolvable experimentally.
Such a scaling behavior is observed here for the first time. Finally, we can
observe how the scaling behavior breaks down for smaller tunnel rates. When
$\Gamma_{N}$ decreases, the $(\Delta\omega_{0})^{2}$ peak versus
$\varepsilon_{d}$ first becomes wider than the $\Delta\Lambda_{0}$ peak (not
shown), before becoming strongly non monotonic (See Figs.3a and 3b).

The remarkable scaling between $\Delta\Lambda_{0}$ and $(\Delta\omega_{0}%
)^{2}$ is directly related to the universality of the AC resistance of a
quantum dot circuit, which was predicted by M. B\"{u}ttiker et al. two decades
ago\cite{Buttiker,Pretre}, and recently revisited as a Korringa Shiba
relation\cite{RC3,RC5,Shiba}. More precisely, for a non-interacting N/dot
junction ($E_{c}=0$) excited at a frequency $\omega_{RF}$ such that
$\Gamma_{N}\gg\omega_{RF},k_{B}T$, the AC resistance is set by $\theta
(\varepsilon_{d})$, i.e. $R_{AC}=h/4e^{2}\theta(\varepsilon_{d})$, which gives
$R_{AC}=h/4e^{2}$ for our spin-degenerate case, for any gate voltage. This
effect can be understood as a quantum charge relaxation effect, which involves
the internal coherent dynamics of the quantum dot. So far, the universality of
the quantum charge relaxation had been observed only with a strongly
spin-polarized GaAs 2-dimensional electron gas device\cite{Gabelli}. Here, we
present the second example of system, i.e. a spin-degenerate carbon nanotube
device, whose behavior is consistent with this phenomenon. Indeed, the scaling
behavior between $\Delta\Lambda_{0}$ and $(\Delta\omega_{0})^{2}$ reveals the
independence of $R_{AC}$ from the dot gate voltage $V_{g}$ (or equivalently
from the dot orbital energy $\varepsilon_{d}$), a property which could not be
probed in Ref.\cite{Gabelli}, and which is already valid for intermediate
tunnel rates.

The fact that we model the charge susceptibility of the quantum dot circuit
with a non-interacting model in spite of Coulomb blockade is non trivial. This
approach is useful to understand our data because we are in a deep Coulomb
blockade regime ($\Gamma_{N}\ll E_{c}$) where correlations effects induced by
interactions (e.g. Kondo effect) are weak. In this limit, one can expect $G$
and $\chi$ to show variations similar to those of the non-interacting case,
with only quantitative modifications. In particular, a reduction of the
amplitude of the signals is expected, due to the reduction of the dot
occupation by Coulomb blockade\cite{Meir}. Our results suggest that in our
regime of parameters, interactions simply lead to a renormalization of our
fitting parameters. In principle, it is possible to generalize our model to
the interacting case to study quantitatively the effects of a finite $E_{c}$
\cite{Meir,Yeyati}. This is beyond the scope of the present article. Anyhow,
our observation of the scaling between $\Delta\Lambda_{0}$ and $(\Delta
\omega_{0})^{2}$ is independent from any theoretical assumption on the dot
interaction regime since the calculation of these quantities from the raw data
only relies on Eq.(\ref{InOut}). The fact that we observe the scaling behavior
between $\Delta\Lambda_{0}$ and $(\Delta\omega_{0})^{2}$ in spite of strong
Coulomb blockade is remarkable. This is in agreement with recent theory works
which suggest that the universality of the charge relaxation resistance
$R_{AC}$ persists in the spin-degenerate interacting case\cite{Minchul,RC5}.

\section{Negative photon damping by a N/dot/S bi-junction\label{HB}}

\begin{figure}[t]
\includegraphics[width=1.\linewidth]{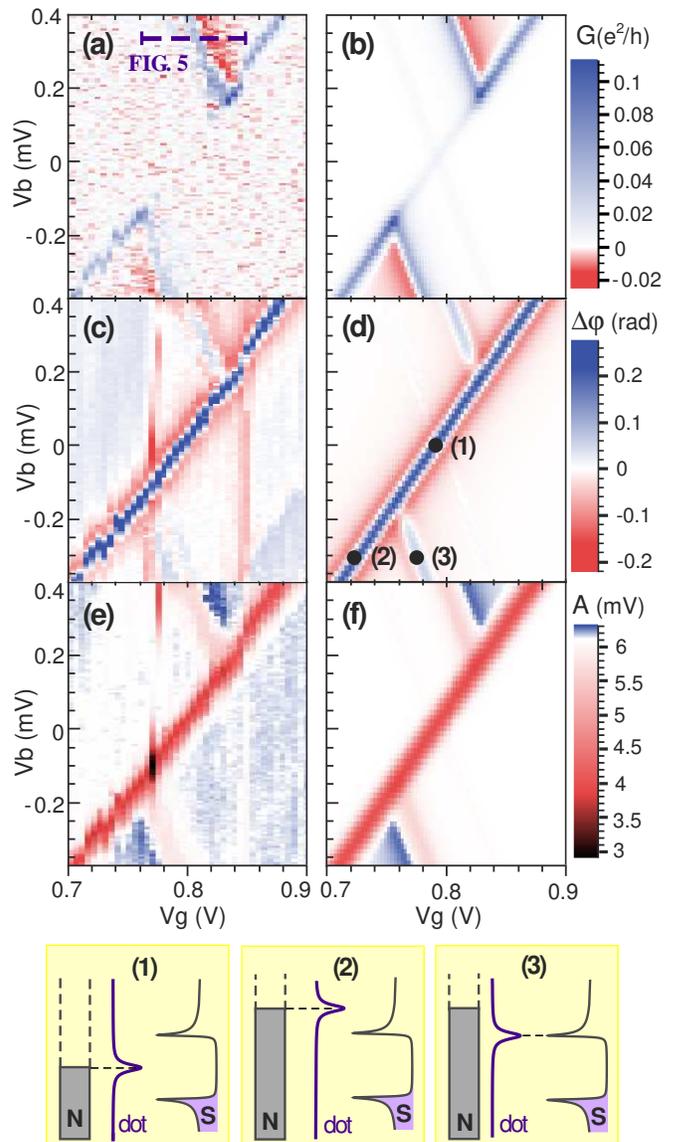}\caption{Panels (a), (c)
and (e): Measured linear conductance $G$, phase shift $\Delta\varphi$ and
total amplitude $A$ of the transmitted microwave signal, versus the dot gate
voltage $V_{g}$ and the bias voltage $V_{b}$. Panels (b), (d) and (d):
Predictions from Eqs. (\ref{Xitot}), (\ref{rr}) and (\ref{II}), for
$\Gamma_{N}/2\pi=0.6~\mathrm{GHz}$, $\Gamma_{S}/2\pi=65~\mathrm{MHz}$,
$\Gamma_{n}/2\pi=8~\mathrm{GHz}$, $g/2\pi=99~\mathrm{MHz}$, $\Delta
=0.17~\mathrm{meV}$, $T=90~\mathrm{mK}$, $\omega_{0}/2\pi=6.65$~\textrm{GHz},
$A_{0}=6.1~\mathrm{mV}$, and $\Lambda_{0}/2\pi=0.259$~\textrm{MHz}$.$ The
white color corresponds to $A=A_{0}$ in panels (e) and (f). Panels (1), (2)
and (3): Electric potential configuration corresponding to the black points in
panel (d).}%
\label{2D}%
\end{figure}

A common belief is that a fermionic reservoir should necessarily damp cavity
photons since it calls for irreversible processes. Is it possible to go
against this natural trend? To answer this question, we consider the finite
bias voltage regime $V_{b}\neq0$ where our device implements a N/dot/S
bi-junction. This can be confirmed from the bi-junction conductance versus
$V_{b}$ and the dot gate voltage $V_{g}$ (Fig. \ref{2D}a). Like in
Fig.\ref{setup}d, we observe two Coulomb triangles which do not close on the
$V_{b}=0$ line but at $eV_{b}\sim\pm\Delta$, and which are shifted along the
$V_{g}$ axis. These features are typical of a N/dot/S structure and are due to
the gap and BCS peaks in the density of states of the S
contact\cite{Dirks,Pfaller,Gramich}. The conductance resonances corresponding
to an alignment between the dot level and the BCS peaks display negative
differential resistance areas\cite{Pfaller} (see red areas in Fig. \ref{2D}a).
This can be understood easily in the limit $\Gamma_{S}\ll k_{B}T$, where, from
a Fermi's golden rule argument, the conductance is proportional to the
derivative of the BCS peak\cite{Ralph}. It is also interesting to notice that
the conductance above the gap has a small amplitude $\left\vert G\right\vert
<0.12\times e^{2}/h$, which suggests a strong asymmetry between the tunnel
rates $\Gamma_{N}$ and $\Gamma_{S}$ to the $N$ and $S$ contacts. A theoretical
modeling of the conductance with Eq.(\ref{II}) of Appendix B confirms that for
the dot level considered in this section, one has $\Gamma_{S}\ll\Gamma
_{N}<k_{B}T$ (see Fig.\ref{2D}b).

We have measured the cavity signals simultaneously with $G$ (Figs. \ref{2D}c
and \ref{2D}e). In agreement with section \ref{LB}, $\Delta\varphi$ and
$\Delta A$ reveal the resonance between the dot level and the Fermi energy of
the N contact even inside the gap area ($e\left\vert V_{b}\right\vert <\Delta
$), in contrast to what happens for $G$. Sign changes in $\Delta\varphi$
similar to those of Fig.\ref{MosaikLittle.eps}d indicate that we are in a
regime with $\Gamma_{S},\Gamma_{N}\ll\omega_{0}$. The microwave amplitude $A$
shows a more surprising behavior. Indeed, the resonances of the dot level with
S and N do not affect similarly the $A$ signal. For $e\left\vert
V_{b}\right\vert >\Delta$, the resonances with the S contact are closely
followed by an area with $\Delta A>0$, which indicates a counterintuitive
negative photon damping (or photon emission) caused by a fermionic reservoir
(see dark blue areas in Fig.\ref{2D}e). So far, with quantum dots circuits
coupled to cavities, photon emission had been obtained only due to tunneling
between two discrete dot levels\cite{Liu,Liu2,Stockklauser,Liu3}.

To model the cavity response, we use again Eqs.(\ref{Xitot}) and (\ref{rr}),
with expressions of $\mathcal{\check{G}}^{r/a}(\omega)$ and $\check{\Sigma
}^{<}(\omega)$ which take into account the finite $\Gamma_{S}$ (see
Eqs.(\ref{green})-(\ref{DD}) of Appendix B). We can reproduce quantitatively
the three signals $\Delta\varphi$, $\Delta A$ and $G$ versus $V_{b}$ and
$V_{g}$ with a consistent set of parameters (see Figs. \ref{2D}b, \ref{2D}d
and \ref{2D}f). The good agreement between the data and theory is also visible
in Fig.~\ref{coupeV} of Appendix A for constant values of $V_{b}$. In
particular, our theory reproduces well the positive $\Delta A$ areas. We take
into account the lever arms determining the shift of the dot and reservoir
energy levels with $V_{b}$ and $V_{g}$. We also use the gap value
$\Delta=0.17~\mathrm{meV}$ given straightforwardly by the $G(V_{b},V_{g})$
map\cite{Gramich}. Then, there remains only 5 adjustable parameters:
$\Gamma_{N}$, $\Gamma_{S}$, $g$, $T$, and the broadening parameter $\Gamma
_{n}$ for the BCS peaks. Fitting simultaneously three two-dimensional plots in
these conditions is non-trivial and possible only due to the adequacy of our
model. The agreement with the data is optimal for $\Gamma_{N}/2\pi
=0.6~\mathrm{GHz}$, $\Gamma_{S}/2\pi=65~\mathrm{MHz}$, $\Gamma_{n}%
/2\pi=8~\mathrm{GHz}$, $g/2\pi=99~\mathrm{MHz}$, and $T=90~\mathrm{mK}$. The
Eq.(\ref{II}) used to model $G$ has been obtained in the absence of the cavity
($g=0$). This approximation is relevant because the cavity brings only small
corrections to this expression, not resolvable in our experiment. In contrast,
$\Delta\varphi$ and $A$ are calculated to second order in $g$. We have again
used a non-interacting approach to model the dot behavior. This approximation
is relevant to understand our data because we are in the deep Coulomb blockade
regime and because Andreev reflections (which are very sensitive to
interactions) are negligible in the small $\Gamma_{S}$ limit. Therefore,
interactions should only induce quantitative modifications of the dot/lead
resonances. Note that a temperature $T=60~\mathrm{mK}$ is optimal to interpret
the low bias voltage data of section \ref{LB}, but we need a higher
temperature $T=90~\mathrm{mK}$ to interpret the finite bias voltage data of
section \ref{HB}. This may be due to heating effects caused by $V_{b}\neq0$,
or to interactions which can modify the dot occupation and thus the amplitude
of dot/lead resonances in the out-of-equilibrium regime.

\begin{figure}[t]
\includegraphics[width=1.\linewidth]{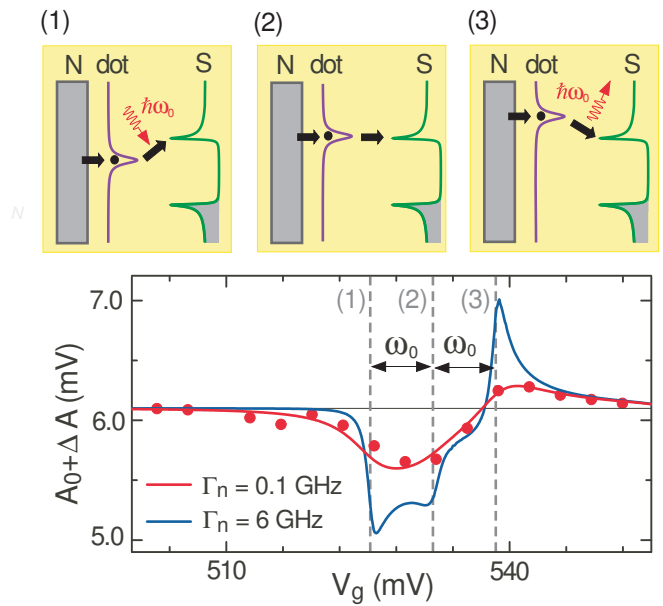}\caption{Conductance $G$ and
microwave amplitude $A$ versus $V_{g}$ (red dots), measured along the dashed
line in Fig.\ref{2D}a for $V_{b}=0.336~\mathrm{mV}$, and theory using the same
parameters as in Fig.\ref{2D} and $\Gamma_{n}/2\pi=8~\mathrm{GHz}$ (red lines)
or $\Gamma_{n}/2\pi=1~\mathrm{GHz}$ (blue lines). The theoretical $G$ for
$\Gamma_{n}/2\pi=1~\mathrm{GHz}$ has been multiplied by $0.2$. Panels (1), (2)
and (3) illustrate the transport regimes corresponding to the gray dashed
lines. In panel (2), the dot orbital is resonant with a BCS peak in the DOS of
the S reservoir. In panels (1)/(3), an electron can pass from the dot orbital
to the BCS peak by absorbing/emitting a cavity photon.}%
\label{pat}%
\end{figure}

Are the $G<0$ and $\Delta A>0$ effects related? In order to answer this
question, Fig.~\ref{pat} shows the measured $A$ and $G$ versus $V_{g}$ (red
dots) together with the theory of Fig.\ref{2D} (red lines), for a constant
bias voltage $V_{b}=0.336~\mathrm{mV}$, along the dashed line in
Fig.\ref{2D}a. These signals vary smoothly due to the large value of
$\Gamma_{n}$. It is very instructive to use a smaller BCS peak broadening
parameter $\Gamma_{n}/2\pi=1~\mathrm{GHz}$ for the theory (blue lines). The
$A$ signal then shows a cusp when the dot level is at resonance with a BCS
peak (gray dashed line (2)) or shifted by $\pm\hbar\omega_{0}/\alpha$ (gray
dashed lines (1) and (3)), with $\alpha$ the lever arm associated to $V_{g}$.
This indicates inelastic tunneling accompanied by photon absorption or
emission along lines (1) and (3). More precisely, in the configuration
corresponding to panel (1)/(3) of Fig.~\ref{pat}, the BCS peaks of the S
contact reinforce the probability of photon absorption/emission, leading to a
pronounced negative/positive $\Delta A$ peak. In contrast, one keeps $\Delta
A<0$ near the N/dot resonance because the density of states of the $N$ contact
can be considered as constant. As expected, the theoretical $G$ for
$\Gamma_{n}/2\pi=1~\mathrm{GHz}$ and $g=0$ does not show cusps along lines (1)
and (3) since this quantity does not take into account photon emission or
absorption (see blue line in top panel of Fig.~\ref{pat}). Since the
experimental $G$ is dominated by the zeroth order contribution in $g$, it
implies that the $G<0$ and $\Delta A>0$ effects in our data are not directly
related. It is more correct to state that these two effects have a common
origin. More precisely, $G<0$ is due to the fact that the DOS of S decreases
with energy in certain areas, which leads to a reduction of the dot current,
whereas $\Delta A>0$ is due to the strong DOS peaks which reinforce photon
emission. Note that Fig.\ref{setup} shows extra resonant lines parallel to the
Coulomb diamond borders, which can be attributed to excited states of the
quantum dot. The excitation energy $E_{ex}$ of these levels is such that
$E_{ex}\gtrsim0.30~\mathrm{meV}\gg\hbar\omega0\simeq0.027~$\textrm{meV.}%
\textbf{ }Furthermore, the measurements of section \ref{HB} have been realized
in another gate voltage range where such excited states are not visible.
Hence, the dot excited states can be disregarded to discuss photon-assisted
tunneling. 

It is important to replace the above results in a wider context.
Photon-assisted tunneling has been observed for 40 years in SIS junctions
\cite{Tien} and for 25 years in quantum dots with N
contacts\cite{Kouwenhoven1,Kouwenhoven2,Blick}. In these seminal experiments,
a broad band coupling scheme was used instead of a cavity and the
photo-induced current was directly measured. In this context, one novelty of
our work is that we use a highly resonant microwave technique to inject and
probe photons. We send a limited photonic power on the quantum dot circuit, so
that the photo-induced current is not resolvable. More precisely, along lines
(1) and (3), the rate of photon absorption/emission ($\Gamma_{e/a}$) by the
quantum dot circuit is $\Gamma_{e/a}\simeq2n_{ph}\Lambda_{0}\Delta A/A_{0}$,
with an average photon number $n_{ph}\sim120$ in the cavity\cite{just}. This
gives $\Gamma_{e/a}\sim2MHz$, which corresponds to a photon-assisted current
of the order of $0.3~\mathrm{pA}$. For comparison, in Ref.\cite{Kouwenhoven1},
the photon-assisted current between a dot and a N contact reaches
$30~\mathrm{pA}$. In spite of this, we can directly detect photon
emission/absorption thanks to the cavity. This demonstrates that circuit QED
techniques provide accurate tools to revisit the physics of photon-assisted
tunneling. Note that despite long-standing theoretical
interest\cite{Zhao,Sun2,Nurbawono,Whan,Cho,Nguyen,
Zhu,Moghaddam,Avriller,Baranski}, our work represents the first experimental
study of photon-assisted tunneling between a quantum dot and a superconductor.

\section{Summary, extension of our theory and perspectives\label{di}}

We have studied experimentally the behavior of a spin-degenerate N/dot/S
hybrid structure based on a carbon nanotube, coupled to a microwave cavity
with frequency $\omega_{0}$. We have observed a large variety of effects
depending on the values of the tunnel rates and on the bias voltage applied to
the device. For intermediate N/dot tunnel rates $\Gamma_{N}\sim2.5\omega_{0}$
and equilibrium conditions, the cavity frequency and linewidth shifts follow a
scaling relation which is independent of the quantum dot gate voltage. This
behavior is related to the universality of the quantum charge relaxation
resistance $R_{AC}$ predicted by B\"{u}ttiker and
coworkers\cite{Buttiker,Pretre} in the adiabatic limit ($\omega_{0}\ll
\Gamma_{N}$). More precisely, it reveals the independence of $R_{AC}$ from the
dot gate voltage, which is already approximately valid for intermediate tunnel
rates $\Gamma_{N}\sim2.5\omega_{0}$. Remarkably, we obtain this behavior in
spite of the presence of Coulomb blockade in the dot, which was not taken into
account in the original model by B\"{u}ttiker et al. This observation is
consistent with recent theory works which predict that the universality of
charge relaxation should persist in the spin-degenerate interacting
case\cite{Minchul,RC5}. Our measurements are doubly complementary to those of
Gabelli et al., who have observed the universal charge relaxation in the
(spin-polarized) non-interacting case, and who could not probe the gate
dependence of $R_{AC}$\cite{Gabelli}. We have also observed in a controlled
way the departure from the scaling regime, when the dot behavior changes from
capacitive to inductive. In the finite bias voltage regime, we have observed
negative photon damping by the quantum dot circuit. This reveals photon
emission caused by inelastic quasiparticle tunneling between the dot and the
BCS peaks of the S reservoir. The cavity signals are able to reveal this
process although it is not resolvable in the dot DC current. Strikingly, all
the effects depicted above can be modeled quantitatively with a single
non-interacting description. Hence, in our regime of parameters, strong
Coulomb blockade, which we have disregarded, does not seem to modify the main
physical behavior of our system. The agreement between our data and theory
suggests that interactions simply lead to a renormalization of our fitting
parameters. Nevertheless, a comparison between our data and a fully
interacting theory would be interesting. So far, theory works have mainly
focused on the value of $R_{AC}$ for $V_{b}=0$ but our work shows that the
cavity frequency shift and linewidth shift would deserve to be studied
independently in the full $V_{g}$ and $V_{b}$ ranges. More generally, our work
validates a description of mesoscopic QED experiments in terms of an
electronic charge susceptibility.

Considering the agreement of our theory with experimental data, it is
interesting to generalize it to more complex hybrid structures. The
versatility of nanofabrication techniques allows to envision a large variety
of experiments combining quantum dot circuits and cavities. In practice,
nanoconductors can be tunnel-coupled to various types of fermionic reservoirs
such as normal metals, superconductors\cite{DeFranceschi}, but also
ferromagnets with collinear\cite{Cottet,CPF} or non collinear
magnetizations\cite{Viennot2,Dora}. These different elements can be combined
in a large variety of geometries, involving for instance inter-dot
hopping\cite{Petersson,Schroer,Frey2,Toida1,Basset,Zhang,Viennot,Liu,Liu2,Stockklauser,Liu3,Viennot2}%
, and multiterminal contacting\cite{Leturcq,CPF}. In this context, we
generalize our approach to geometries with several quantum dots/sites or
several orbitals. In the case where each discrete level $i\in\lbrack1,N]$ of
the nanocircuit is shifted by the cavity field $\hat{a}+\hat{a}^{\dag}$ with a
constant $g_{i}$, we obtain%

\begin{equation}
\frac{b_{t}}{b_{in}}=\frac{t_{0}}{\omega_{RF}-\omega_{0}-i\Lambda_{0}%
-\Sigma_{i,j}g_{i}g_{j}\chi_{ij}(\omega_{0})} \label{tt}%
\end{equation}
In the linear response limit, the susceptibility $\chi_{ij}(\omega)$ for
orbital indices $i,j$ can be calculated within the Keldysh formalism
as\cite{check}%
\begin{equation}
\chi_{i,j}^{\ast}(\omega_{0})=-i%
%TCIMACRO{\tint }%
%BeginExpansion
{\textstyle\int}
%EndExpansion
\frac{d\omega}{2\pi}\mathrm{Tr}\left[  \mathcal{\check{S}}_{ij}(\omega
)\mathcal{\check{G}}^{r}(\omega)\check{\Sigma}^{<}(\omega)\mathcal{\check{G}%
}^{a}(\omega)\right]  \label{Xigg}%
\end{equation}
with%
\begin{equation}
\mathcal{\check{S}}_{ij}(\omega)=\check{\tau}_{i}\mathcal{\check{G}}%
^{r}(\omega+\omega_{0})\check{\tau}_{j}+\check{\tau}_{j}\mathcal{\check{G}%
}^{a}(\omega-\omega_{0})\check{\tau}_{i} \label{CC}%
\end{equation}
These expressions involve multisite Keldysh Green's functions $\mathcal{\check
{G}}^{r(a)}$, a lesser self energy $\check{\Sigma}^{<}$ and the
electron/photon coupling element $\check{\tau}_{i}$ at site $i$, which are
defined in Appendix \ref{multisite}. In principle, the susceptibility
$\chi_{ij}(\omega)$ can be calculated with other techniques than the Keldysh
formalism\cite{defXi,example}. However, one interest of this approach is that
it is particularly convenient for describing non equilibrium configurations
with multiple quantum dots and multiple reservoirs (normal metals,
ferromagnets, and superconductors), as illustrated for instance by Ref.
\cite{Trocha}. It goes beyond the sequential tunneling picture used so far to
interpret most Mesoscopic QED experiments. \ In principle, it also enables the
description of Coulomb interactions\cite{Meir,Yeyati}.

The above formalism could be instrumental for understanding the behavior of
complex cavity/nanocircuit hybrid structures. In particular, it is suitable
for understanding the interaction between cavity photons and Cooper pair
splitters, or topological hybrid nanocircuits. Non-local entanglement and
self-adjoint Majorana bound states are intensively sought after in these
devices, and new investigation tools such as cavity photons could be
instrumental in this quest. More generally, our results show that mesoscopic
QED represents a powerful toolbox to investigate quantum charge relaxation,
photon-assisted tunneling, and all other effects involving tunneling between a
discrete level and fermionic reservoirs. This opens many possibilities. For
instance, the dynamics of the many body Kondo effect could be explored thanks
to circuit QED techniques. Quantum dot circuits could also open new
possibilities for the so-called "quantum reservoir
engineering"\cite{Sarlette,Holland}, which would exploit fermionic reservoirs
in non-equilibrium configurations to prepare non trivial photonic and
electronic states. Finally, there is a direct analogy between our setup and a
quantum dot circuit coupled to the vibrational modes of a
nano-object\cite{Steele}. Hence, our findings could be transposed to
understand the dissipation of nano electro mechanical systems.

\section{Acknowledgements}

We acknowledge useful discussions with M. B\"{u}ttiker, T. Cubaynes, R.
Deblock, G. F\`{e}ve and F. Mallet. This work was financed by the ERC Starting
grant CirQys, the EU FP7 project SE2ND[271554], and the ANRNanoQuartet
[ANR12BS1000701] (France).\begin{figure}[t]
\includegraphics[width=1\linewidth]{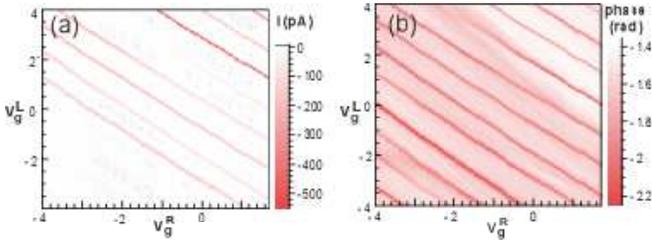}\caption{Current $I$ through
the quantum dot (left panel) and cavity signal $\Delta\varphi$ (right panel)
versus $V_{g}^{L}$ and $V_{g}^{R}$, in the area corresponding to the data of
Figs.\ref{ratio}c and \ref{ratio}d.}%
\label{SingleDot.eps}%
\end{figure}

\section{Appendix A: Experimental details}

\subsection{Sample fabrication and control\label{sample2}}

The cavity is a half-wavelength superconducting Nb transmission line
resonator, with a bare frequency $\omega_{0}/2\pi\sim6.65$~\textrm{GHz} and a
quality factor $Q\sim12800$. We measure the cavity transmission $b_{t}/b_{in}$
with a homodyne-like detection scheme. To form the quantum dot circuit, we use
a carbon nanotube grown by chemical vapor deposition, which is stamped into
the cavity to preserve $Q$\cite{Viennot0}. We evaporate on the nanotube two N
contacts formed by 70$~\mathrm{nm}$ of Pd, and a S contact formed by
4$~\mathrm{nm}$ of Pd proximized with 100$~\mathrm{nm}$ of Al. The nanotube
sections on the left and right of the S contact are coupled to remote DC gates
with voltages $V_{g}^{L}$ and $V_{g}^{R}$ (see Fig.\ref{setup}a). The AC top
gate TG consists of a trilayer \textrm{Al}$_{2}$\textrm{O}$_{3}$
(6nm)/Al(50nm)/Pd(20nm). The double dot design\ of our sample was initially
developed for a Cooper pair splitting experiment which will be reported
elsewhere\cite{CPS}.

\begin{figure}[t]
\includegraphics[width=1\linewidth]{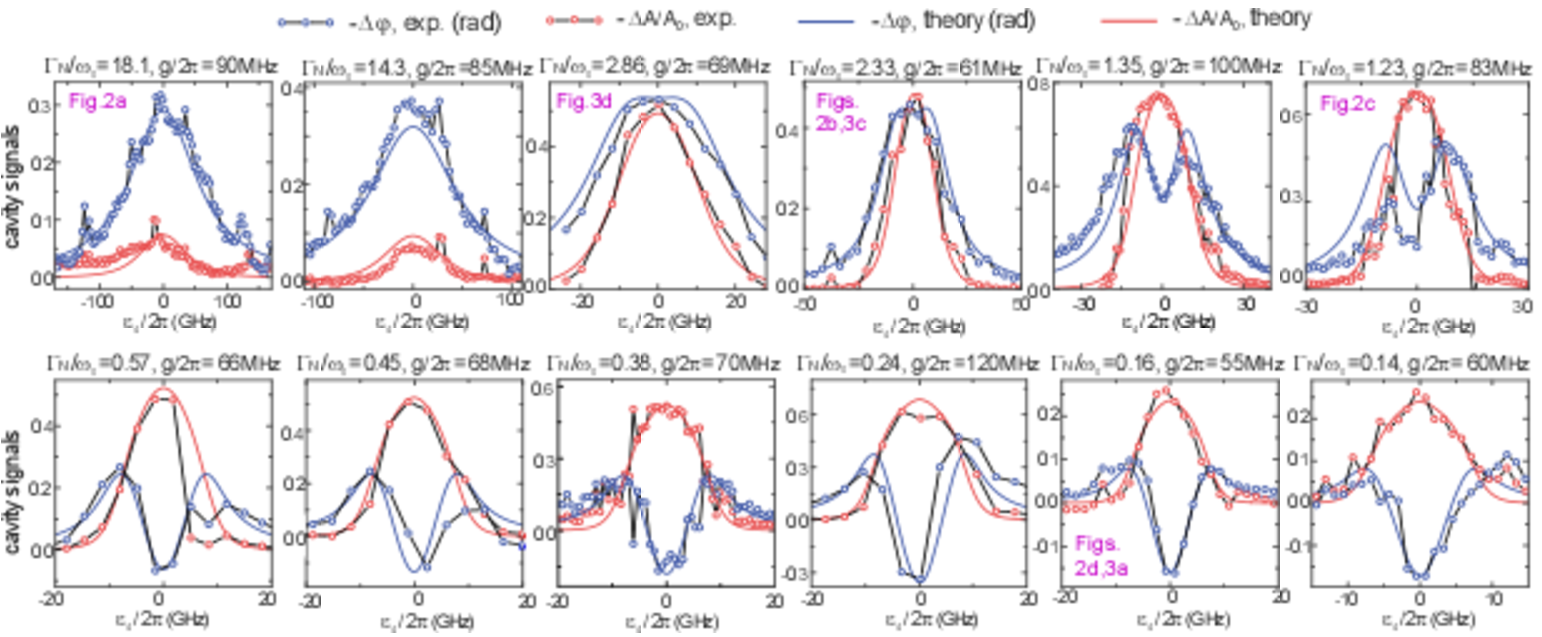}\newline\newline%
\begin{tabular}
[c]{|l|}\hline
measurement shot\\\hline
$V_{g}^{0}$ (V)\\\hline
$\alpha/2\pi$ (GHz.V$^{-1}$)\\\hline
$g/2\pi$ (MHz)\\\hline
$\Gamma_{N}/\omega_{0}$\\\hline
$\Gamma_{N}/2\pi$ (GHz)\\\hline
$\widetilde{\Gamma_{N}}/2\pi$ (GHz)\\\hline
$\widetilde{V_{b}}$ (mV)\\\hline
\end{tabular}
~%
\begin{tabular}
[c]{|c|c|}\hline
$1$ & $1$\\\hline
$-0.88$ & $0.09$\\\hline
$200$ & $200$\\\hline
$69$ & $61$\\\hline
$2.86$ & $2.33$\\\hline
$19$ & $15.5$\\\hline
$21$ & $16$\\\hline
$-0.18$ & $-0.37$\\\hline
\end{tabular}
~%
\begin{tabular}
[c]{|c|}\hline
$2$\\\hline
$0.67$\\\hline
$667$\\\hline
$55$\\\hline
$0.16$\\\hline
$1.05$\\\hline
$\times$\\\hline
$\times$\\\hline
\end{tabular}
~%
\begin{tabular}
[c]{|c|c|c|}\hline
$3$ & $3$ & $3$\\\hline
$-1.30$ & $-0.65$ & $0.47$\\\hline
$629$ & $629$ & $629$\\\hline
$100$ & $83$ & $60$\\\hline
$1.35$ & $1.23$ & $0.14$\\\hline
$9.00$ & $8.20$ & $0.9$\\\hline
$12.5$ & $7.5$ & $\times$\\\hline
$0.42$ & $0.32$ & $\times$\\\hline
\end{tabular}
~%
\begin{tabular}
[c]{|c|}\hline
$4$\\\hline
$0.75$\\\hline
$657$\\\hline
$66$\\\hline
$0.57$\\\hline
$3.8$\\\hline
$\times$\\\hline
$\times$\\\hline
\end{tabular}
~%
\begin{tabular}
[c]{|c|}\hline
$5$\\\hline
$1.16$\\\hline
$765$\\\hline
$70$\\\hline
$0.38$\\\hline
$2.5$\\\hline
$\times$\\\hline
$\times$\\\hline
\end{tabular}
~%
\begin{tabular}
[c]{|c|c|c|c|}\hline
$6$ & $6$ & $6$ & $6$\\\hline
$-2.03$ & $-1.41$ & $-0.72$ & $-0.16$\\\hline
$704$ & $704$ & $704$ & $704$\\\hline
$68$ & $120$ & $85$ & $90$\\\hline
$0.45$ & $0.24$ & $14.3$ & $18.1$\\\hline
$3$ & $1.6$ & $95$ & $120$\\\hline
$\times$ & $\times$ & $64$ & $149$\\\hline
$\times$ & $\times$ & $0.24$ & $0.24$\\\hline
\end{tabular}
\caption{Top panels: Cavity signals $\Delta\varphi$ (blue dots) and $\Delta
A/A_{0}$ (red dots) versus $\varepsilon_{d}$ for $V_{b}=0$ and different dot
orbitals with decreasing tunnel rates $\Gamma_{N}$ \ from top to bottom and
left to right panels. The red and blue lines show the predictions given by
Eqs.(\ref{XiT}) and (\ref{rr}) for the values of $\Gamma_{N}$ and $g$ given in
the different panels and $T=60~\mathrm{mK}$. When a resonance is already shown
in the main text, we indicate the corresponding figure number in pink. Bottom
table: Parameters corresponding to the different N/dot resonances shown in the
upper part of the figure. We first show the effective gate voltage $V_{g}^{0}$
and the lever arm $\alpha$ extracted from our experimental data, and the
fitting parameters $\Gamma_{N}$ and $g$ used to model the cavity signals
$\Delta\varphi$ and $\Delta A$. We also show, when possible, the value of the
N/dot tunnel rate $\widetilde{\Gamma_{N}}$ estimated from the conductance data
through the S/dot/N structure for a voltage $\widetilde{V_{b}}>\Delta/e$. Each
block in the table corresponds to the one shot measurement of a given gate
voltage range. The signals $\Delta\varphi$, $\Delta A$ and $G$ were measured
simultaneously in each measurement shot.}%
\label{MosaikBig}%
\end{figure}

The sample is placed in a dilution refrigerator with a base temperature of
$16~$\textrm{mK}. We apply the same bias voltage $V_{b}$ to the two N
contacts. We measure the current $I$ in the S contact with a DC measurement,
and we use a Lock-In detection to determine the corresponding differential
conductance $G$. For $e\left\vert V_{b}\right\vert >\Delta$, the current $I$
versus $V_{g}^{L}$ and $V_{g}^{R}$ and the cavity signals correspond to a
pattern of parallel lines (see Fig.\ref{SingleDot.eps}). This indicates that
the whole nanotube section between the two N contacts behaves as a single
quantum dot. Therefore, in section \ref{HB}, we use an effective gate voltage
parameter $V_{g}=a_{L}V_{g}^{L}+a_{R}V_{g}^{R}$ to represent the data. In
section \ref{LB}, we use level-dependent lever arms $\alpha$ and gate voltage
offsets $V_{g}^{0}$ to express the gate voltage axis in terms of the energy
$\varepsilon_{d}=\alpha(V_{g}-V_{g}^{0})$ of the considered dot level with
respect to the Fermi energy of the N reservoir (see Table of
Fig.\ref{MosaikBig} for the values of the parameters $\alpha$ and $V_{g}^{0}%
$). This is more convenient to compare the energy width of the different
resonances.

\subsection{Supplementary data and system parameters}

\begin{figure}[t]
\includegraphics[width=0.7\linewidth]{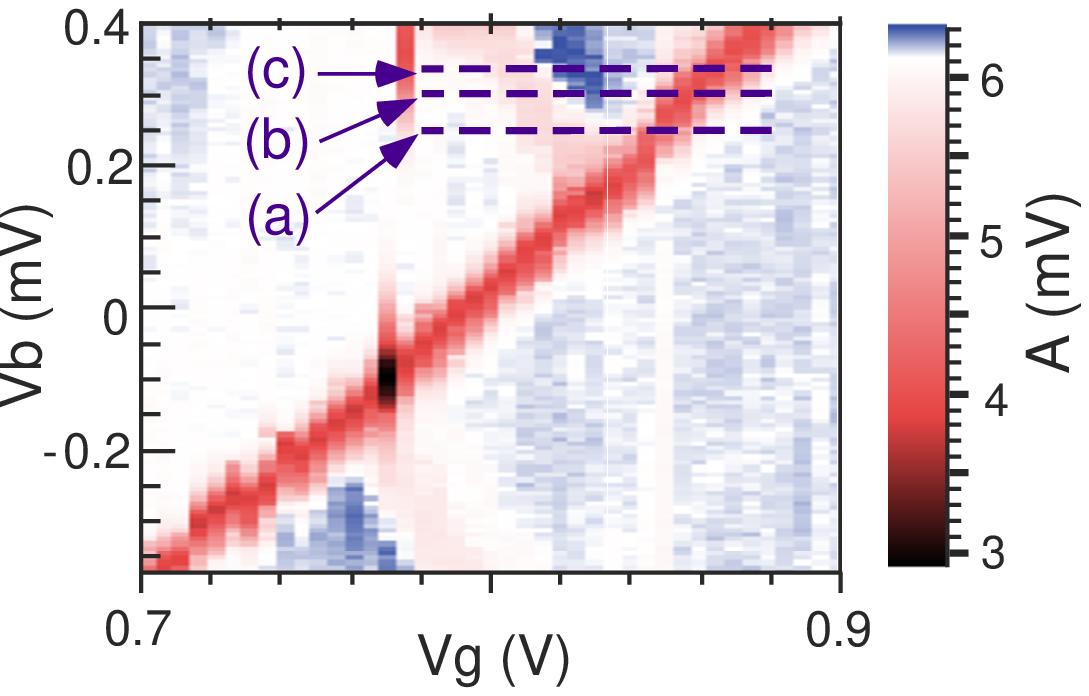}
\includegraphics[width=1.\linewidth]{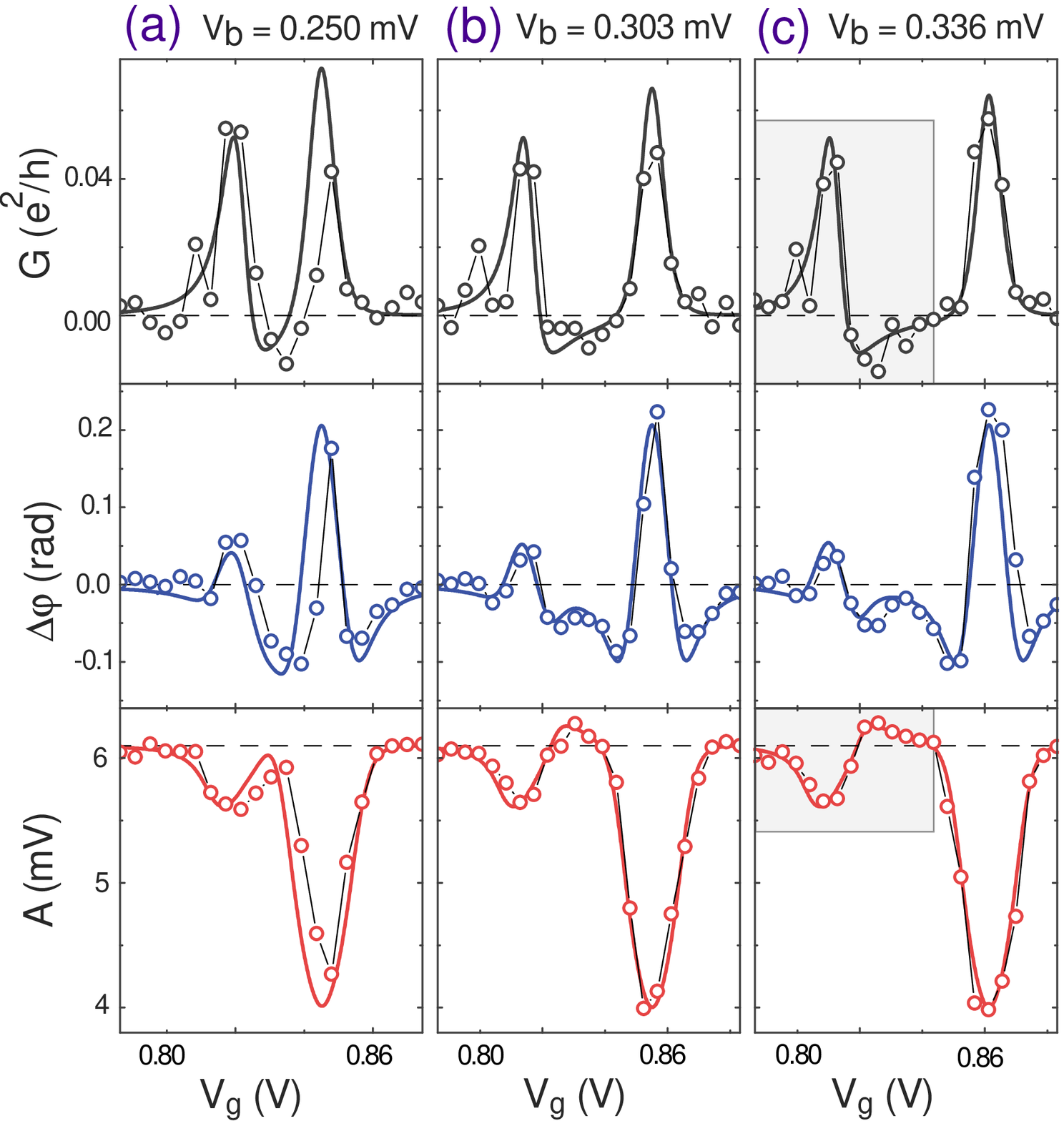}\caption{Top panel:
Measured amplitude $A$ versus $V_{b}$ and $V_{g}$, already shown in
Fig.\ref{2D},c. Bottom panels: Measured conductance $G$ (black dots), and
cavity signals $\Delta\varphi$ (blue dots) and $A$ (red dots) versus $V_{g}$,
along the dashed lines in the top panel, for $V_{b}=0.25$, $0.303$, and
$0.336~\mathrm{mV}$ from left to right. The full red, blue and black lines
show the predictions given by Eqs. (\ref{Xitot}), (\ref{rr}) and (\ref{II}),
for the same parameters as in Fig.\ref{2D}. The areas in the gray rectancles
are enlarged in Fig. \ref{pat} of the main text.}%
\label{coupeV}%
\end{figure}

In order to demonstrate further the quantitative agreement between our
theoretical approach and the data, we present supplementary data together with
their theoretical modelling. Figure \ref{MosaikBig} shows the cavity signals
$\Delta\varphi$ and $\Delta A$ at $V_{b}=0$ for 12 different quantum
dot/reservoir resonances, including those of Figs.\ref{MosaikLittle.eps} and
\ref{ratio} for completeness. Figure \ref{coupeV} shows the dot conductance
and cavity signals, for different values of $V_{b}$, on a wider $V_{g}$-scale
than in Fig.\ref{pat}.

Near each dot/reservoir resonance, we calibrate the bare cavity linewidth
$\Lambda_{0}\sim2\pi\times0.26~\mathrm{MHz}$ and the bare cavity transmission
amplitude $A_{0}\sim6.1~\mathrm{mV}$. The average photon number $n_{ph}$ in
our measurements is estimated from setup transmission calibration. Assuming a
6dB uncertainty we obtain a lower bound $n_{ph}>20$ which ensures the validity
of the semiclassical approximation used in our theory (see Appendix B). The
agreement between our theory and data also confirms that we remain in the
linear response regime invoked in Appendix B. Otherwise, the width of the
resonances in the cavity response would not match with the theory\cite{tbe}.

The parameters $V_{g}^{0}$, $\alpha$, $\Gamma_{N}$ and $g$ for the 12
resonances\ presented in Figs.\ref{MosaikLittle.eps} and \ref{MosaikBig} are
given in the bottom table of Fig.\ref{MosaikBig}. The dot/photon coupling $g$
varies from $2\pi\times55~\mathrm{MHz}$ to $2\pi\times120~\mathrm{MHz}$ and
the tunnel rate $\Gamma_{N}$ from $2\pi\times0.9~\mathrm{GHz}$ to $2\pi
\times120~\mathrm{GHz}$. The circuit parameters may take different values for
different measurement shots, probably due to charge reorganizations in the
sample, which change the offset $V_{g}^{0}$ or the potential landscape of the
quantum dot. Therefore, we have separated the table in Fig.\ref{MosaikBig}
into different blocks which correspond to single shot measurements of a given
gate voltage range. The tunnel rate $\Gamma_{N}$ does not show a monotonic
dependence with $V_{g}$ on a large scale. Sometimes, we have found a locally
monotonic dependence, on a scale of about 3 consecutive resonances, as
illustrated by the blocks corresponding to measurements 3 and 6 in the table.
The non-monotonic behavior of $\Gamma_{N}$ with $V_{g}$ is very common in
carbon nanotubes and may be attributed to weak disorder. Moreover, the
variations of $\Gamma_{N}$ and $g$ do not seem correlated, most probably
because $\Gamma_{N}$ depends on the properties of the dot interfaces whereas
$g$ depends on the overlap of the whole dot orbital with the cavity photonic
pseudopotential\cite{Cottet2015}. Finally, the value of $\alpha$ for the
measurement 1 differs significantly from the values used in the other
measurements because $a_{L}=1$ and $a_{L}=0$ were used for measurement 1
whereas $a_{L}=0.75$ and $a_{R}=0.66$ were used for the other measurements.
For section \ref{HB}, we have used in the theory $\varepsilon_{d}=\alpha
(V_{g}-V_{g}^{0})+\gamma V_{b}$ with $V_{g}^{0}=0.79~$\textrm{V}, $\alpha
=2\pi\times723~$\textrm{GHz.V}$^{-1}$ and $\gamma=2\pi\times87.3~$%
\textrm{GHz.mV}$^{-1}$.

Since we are in the regime $\Gamma_{N}\gg\Gamma_{S}$, we could expect that,
for each dot orbital considered in section III, a fit of the N/dot conductance
peak for a bias voltage $\widetilde{V_{b}}>\Delta/e$ should give the value of
the N/dot tunnel rate. Then, the only remaining fitting parameter for
$\Delta\varphi$ and $\Delta A$ should be $g$. Indeed, for the 6 upper
resonances in Fig.\ref{MosaikBig}, a Lorentzian fit of the N/dot conductance
peak gives an estimate $\widetilde{\Gamma_{N}}$ of the N/dot tunnel rate which
is in rather good agreement with the value $\Gamma_{N}$ estimated from the
cavity signals (see values in the bottom table of Fig.\ref{MosaikBig}). For
the highest tunnel rates $\Gamma_{N}/\omega_{0}=18.1$ and $14.3$, inaccuracies
in the estimation of $\widetilde{\Gamma_{N}}$ stem from cotunneling peaks
which appear between the Coulomb diamonds and would require a more complete
theory. The conductance data for intermediate tunnel rates $1.23<\Gamma
_{N}/\omega_{0}<2.86$ are more affected by experimental noise. One can try to
minimize these two difficulties by estimating $\widetilde{\Gamma_{N}}$ for
values $\widetilde{V_{b}}$ of the bias voltage such that cotunneling and
experimental noise are reduced. However, for smaller tunnel rates comparable
to the temperature, the straightforward estimation of the N/dot tunnel rate
from $G$ is not possible anymore due to temperature broadening effects.
Therefore, in section III, we have preferred to treat $\Gamma_{N}$ as a
fitting parameter for the cavity response, knowing that since we have to fit
simultaneously two 1D curves with two parameters $\Gamma_{N}$ and $g$, these
parameters are strongly constrained anyway.

\section{Appendix B: Theoretical approach}

\subsection{Hamiltonian of the quantum dot circuit}

To model the behavior of our setup, we use the total Hamiltonian (\ref{Htot})
of the main text, with%
\begin{align}
H_{d}  &  =\sum\limits_{\sigma}\varepsilon_{d}\hat{d}_{\sigma}^{\dag}\hat
{d}_{\sigma}+\Delta\sum\limits_{k}\left(  \hat{c}_{k\uparrow}^{S\dag}\hat
{c}_{-k\downarrow}^{S\dag}+H.c.\right) \label{Hfull}\\
&  +\sum\limits_{O\in\{S,N\},k,\sigma}\left(  \varepsilon_{k}^{O}\hat
{c}_{k\sigma}^{O\dag}\hat{c}_{k\sigma}^{O}+(t_{O}\hat{d}_{\sigma}^{\dag}%
\hat{c}_{k\sigma}^{O}+H.c)\right) \nonumber\\
&  +\sum\limits_{k,k^{\prime},\sigma}\left(  \varepsilon_{kk^{\prime}}^{n}%
\hat{b}_{kk^{\prime}\sigma}^{n\dag}\hat{b}_{kk^{\prime}\sigma}^{n}+(t_{n}%
\hat{b}_{kk^{\prime}\sigma}^{n\dag}\hat{c}_{k\sigma}^{S}+H.c.)\right)
\nonumber
\end{align}
the Hamiltonian of a single quantum dot contacted to a N and a S contact.
Above $\hat{d}_{\sigma}^{\dag}$ [$\hat{c}_{k\sigma}^{O\dag}$] creates an
electron with spin $\sigma$ in the orbital with energy $\varepsilon_{d}$
[$\varepsilon_{k}^{O}$] of the dot [reservoir $O\in\{S,N\}$]. To account for
the broadening of the BCS peaks in the density of states of S, we use an
auxiliary reservoir $n$ whose states can be populated by the operators
$\hat{b}_{kk^{\prime}\sigma}^{n\dag}$. For simplicity, each level $k\sigma$ of
$S$ is coupled to an independent set $kk^{\prime}\sigma$ of levels in $n$. We
assume that a bias voltage $V_{b}$ is applied to the N contact whereas the S
contact is grounded. For simplicity, we disregard Coulomb interactions in the
whole Appendix B. Throughout this paper, we use $\hbar=1$ and define the
quantities $\Gamma_{N}$, $\Gamma_{S}$, $\Gamma_{n}$, $\varepsilon_{d}$,
$\omega_{0}$, $\omega_{RF}$, $\Delta\omega_{0}$, $\Delta\Lambda_{0}$ and $g$
as pulsations.

\subsection{Calculation of the cavity microwave transmission\label{smcl}}

The 2-port transmission of the cavity can be calculated with the input-output
formalism for microwave cavities\cite{Devoret}. In this framework, the bosonic
modes $q$ in Eq. (\ref{Htot}) include propagating modes in the $L$ and $R$
ports of the cavity, and extra modes accounting for internal cavity damping.
The L and R ports cause contributions $\Lambda_{L(R)}$ to the bare cavity
linewidth $\Lambda_{0}$, related to the $f_{Q}$ coupling factors and the modes
density (see Ref. \cite{Devoret} for details). One can treat explicitly the
excitation with frequency $\omega_{RF}$ imposed on the cavity through the
incoming mode $Q$ of the $L$ port ($\omega_{RF}=\omega_{Q}$), by adding to the
Hamiltonian (\ref{Htot}) the contribution%
\begin{equation}
H_{RF}=-i(f_{Q}\hat{a}^{\dag}B_{Q}e^{-i\omega_{RF}(t-t_{0})}-f_{Q}^{\ast}%
\hat{a}B_{Q}^{\ast}e^{i\omega_{RF}(t-t_{0})}) \label{hrf}%
\end{equation}
with $t_{0}<t$ an initial time before the interaction of the propagating modes
with the cavity. The term $H_{RF}$ corresponds to a classical input signal
\begin{equation}
b_{in}=B_{Q}f_{Q}e^{-i\omega_{RF}(t-t_{0})}/\sqrt{2\Lambda_{L}} \label{in}%
\end{equation}
in port $L$. Disregarding quantum fluctuations in the input modes of the
cavity, Eqs.(\ref{Htot}) and (\ref{hrf}), lead to%
\begin{equation}
\frac{d}{dt}\hat{a}=-i\omega_{0}\hat{a}-ig\hat{n}-\Lambda_{0}\hat{a}%
-\sqrt{2\Lambda_{L}}b_{in} \label{HH}%
\end{equation}
with $\hat{n}(t)=%
%TCIMACRO{\tsum _{\sigma}}%
%BeginExpansion
{\textstyle\sum_{\sigma}}
%EndExpansion
\hat{d}_{\sigma}^{\dag}\hat{d}_{\sigma}$, while the cavity output signal
writes%
\begin{equation}
\hat{b}_{t}=\sqrt{2\Lambda_{R}}\hat{a} \label{out}%
\end{equation}
If the number of photons in the cavity is larger than $\sim10$, we can use the
semiclassical approximation $\hat{a}\simeq\left\langle \hat{a}\right\rangle $.
In the linear response limit and stationnary regime, $\left\langle \hat
{a}\right\rangle $ has a negligible component in $e^{i\omega_{RF}t}$ provided
the loaded quality factor of the cavity remains good and $\omega_{RF}%
\sim\omega_{0}$. In the framework of Eq.(\ref{Hfull}), one can thus estimate
the time variations of the average number of electrons in the dot from the
linear response to $\hat{a}\simeq\bar{a}e^{-i\omega_{RF}t}$, as%
\begin{equation}
\left\langle \hat{n}\right\rangle (t)=g\tilde{\chi}(\omega_{RF})\bar
{a}e^{-i\omega_{RF}t}+g\tilde{\chi}(-\omega_{RF})\bar{a}^{\ast}e^{i\omega
_{RF}t} \label{nn}%
\end{equation}
This expression involves the dot charge susceptibility%
\begin{equation}
\tilde{\chi}(t)=-i\theta(t)\left\langle \{\hat{n}(t),\hat{n}%
(t=0)\}\right\rangle _{g=0}%
\end{equation}
calculated in the absence of the cavity. Throughout appendix B, we use the
quantum mechanics convention for the Fourier transform, i.e. $\tilde{\chi
}(\omega)=%
%TCIMACRO{\tint \nolimits_{-\infty}^{+\infty}}%
%BeginExpansion
{\textstyle\int\nolimits_{-\infty}^{+\infty}}
%EndExpansion
dt~\tilde{\chi}(t)e^{i\omega t}$. Injecting Eq.(\ref{nn}) into the statistical
average of Eq. (\ref{HH}) and disregarding non resonant terms, we obtain%
\begin{equation}
\bar{a}=\frac{-if_{Q}B_{Q}e^{i\omega_{RF}t_{0}}}{\hbar\omega_{RF}-\hbar
\omega_{0}+i\Lambda_{0}-g^{2}\tilde{\chi}(\omega_{0})} \label{aa}%
\end{equation}
For an agreement with the experimental data, one has to keep in mind that
microwave equipment uses the electrical engineering Fourier transform
convention, which is complex conjugated to the usual quantum mechanics
convention. Hence, combining Eqs.(\ref{in}), (\ref{out}) and (\ref{aa}), with
$b_{t}=\left\langle \hat{b}_{t}\right\rangle $, and making the substitution
$i\rightarrow-i$, we obtain Eq.(\ref{InOut}) of the main text, with
$\chi(\omega)=\tilde{\chi}(\omega)^{\ast}$. Note that
Refs.\cite{Cottet2015,Cottet:11,Schiro,Dmytruk} have presented related linear
response approaches to express the cavity behavior in terms of the charge
susceptibility of the quantum dot.

\subsection{Keldysh description of the quantum dot circuit\label{kel}}

Using the time-dependent Keldysh formalism\cite{Jauho}, we obtain the
expression (\ref{Xitot}) of the main text for the dot charge susceptibility.
Interestingly, Refs.\cite{Skoldberg} and \cite{Schiro} have introduced related
expressions, restricted to the N/dot and Andreev molecular cases respectively.
Equation (\ref{Xitot}) involves the retarded, advanced, and lesser Green's
functions $\mathcal{\check{G}}^{c}$ of the quantum dot, with $c=r,a$ and $<$
respectively, which have the structure%
\begin{equation}
\mathcal{\check{G}}^{c}=\left[
\begin{tabular}
[c]{ll}%
$\mathcal{G}_{\hat{d}_{\uparrow},\hat{d}_{\uparrow}^{\dag}}^{c}$ &
$\mathcal{G}_{\hat{d}_{\uparrow},\hat{d}_{\downarrow}}^{c}$\\
$\mathcal{G}_{\hat{d}_{\downarrow}^{\dag},\hat{d}_{\uparrow}^{\dag}}^{c}$ &
$\mathcal{G}_{\hat{d}_{\downarrow}^{\dag},\hat{d}_{\downarrow}^{c}}^{c}$%
\end{tabular}
\ \right]  \label{green}%
\end{equation}
in Nambu space. For any operators $A$ and $B$, we use $\mathcal{G}_{A,B}%
^{r}(t)=-i\theta(t)\left\langle \{A(t),B(t=0)\}\right\rangle $ and
$\mathcal{G}_{A,B}^{<}(t)=i\left\langle B(t=0)A(t)\right\rangle $. From
Hamiltonian (\ref{Hfull}), one obtains\cite{Sun1,Sun2}:%
\begin{equation}
\mathcal{\check{G}}^{r}(\omega)=\left(  \mathcal{\check{G}}^{a}(\omega
)\right)  ^{\dag}=\left[  \omega\check{1}-\check{E}_{dot}-\check{\Sigma}%
^{r}(\omega)\right]  ^{-1} \label{INV}%
\end{equation}%
\begin{equation}
\mathcal{\check{G}}^{<}(\omega)=\mathcal{\check{G}}^{r}(\omega)\check{\Sigma
}^{<}(\omega)\mathcal{\check{G}}^{a}(\omega)
\end{equation}
with%
\begin{equation}
\check{\Sigma}^{r}(\omega)=-i(\Gamma_{N}/2)\check{1}-i(\Gamma_{S}%
/2)\mathcal{\check{C}}(\omega)
\end{equation}%
\begin{equation}
\check{\Sigma}^{<}(\omega)=i\Gamma_{N}\check{f}_{N}(\omega)+i\Gamma
_{S}f(\omega)\operatorname{Re}[\mathcal{\check{C}}(\omega)]
\end{equation}
Above, we have introduced the diagonal matrices $\check{1}=diag(1,1)$,
$\check{E}_{dot}=diag(\varepsilon_{d},-\varepsilon_{d})$ and $\check{f}%
_{N}(\omega)=diag(f(\omega-eV_{b}),f(\omega+eV_{b}))$. The terms
$\check{\Sigma}^{c}(\omega)$, with $c\in\{r,a,<\}$, describe the effect of the
$N$ and $S$ reservoirs on the quantum dot Green's functions in the large
bandwidth approximation. We use tunnel rates $\Gamma_{r}=2\pi\left\vert
t_{r}\right\vert ^{2}\rho_{r}$ with $\rho_{r}$ the density of states per spin
direction in reservoir $r\in\{S,N,n\}$. For describing electronic correlations
in the superconducting reservoir, we use
\begin{equation}
\mathcal{\check{C}}(\omega)=\left[
\begin{tabular}
[c]{ll}%
$G_{\omega}$ & $F_{\omega}$\\
$F_{\omega}$ & $G_{\omega}$%
\end{tabular}
\ \right]
\end{equation}
with $G_{\omega}=-i(\omega+i\frac{\Gamma_{n}}{2})/D_{\omega}$, $F_{\omega
}=i\Delta/D_{\omega}$ and%
\begin{equation}
D_{\omega}=\sqrt{\Delta^{2}-(\omega+i\frac{\Gamma_{n}}{2})^{2}} \label{DD}%
\end{equation}
The parameter $\Gamma_{n}$ is often omitted (see for instance Refs.
\cite{Sun1,Sun2,Whan,Zhao,Cho,Nurbawono,Kang}), but is it essential to account
for the broadening of the BCS peaks which is observed experimentally.

In the case $V_{b}=0$ and $\Gamma_{S}\ll\Gamma_{N},\Delta$ the effect of the
superconducting contact can be disregarded i.e. $\Gamma_{S}=0$. In this limit,
Eq.(\ref{Xitot}) leads to
\begin{equation}
\chi(\omega)=\frac{\Gamma_{N}}{\pi\omega(i\Gamma_{N}-\omega)}Log[\frac
{4\varepsilon_{d}^{2}+\Gamma_{N}^{2}}{4\varepsilon_{d}^{2}-(2\omega
-i\Gamma_{N})^{2}}] \label{Xi0}%
\end{equation}
for $T=0$ and Eq.(\ref{XiT}) for $T$ finite. In the case $\Gamma_{S}\neq0$, we
evaluate $\chi(\omega)$ numerically from Eqs. (\ref{Xitot}), (\ref{SSS}) and
(\ref{green})-(\ref{DD}).

For completeness, we mention that the DC current through the spin-degenerate
quantum dot can be calculated for $g=0$ as\cite{Sun1}:%

\begin{align}
I  &  =\frac{2e\Gamma_{N}\Gamma_{S}}{h}%
%TCIMACRO{\tint }%
%BeginExpansion
{\textstyle\int}
%EndExpansion
d\omega\left(  f(\omega-eV_{b})-f(\omega)\right)  \left[  \mathcal{\check{G}%
}^{r}\operatorname{Re}[\mathcal{\check{C}}]\mathcal{\check{G}}^{a}\right]
_{11}\nonumber\\
&  +\frac{2e\Gamma_{N}^{2}}{h}%
%TCIMACRO{\tint }%
%BeginExpansion
{\textstyle\int}
%EndExpansion
d\omega(f(\omega-eV_{b})-f(\omega+eV_{b}))\left\vert \mathcal{\check{G}}%
_{12}^{r}\right\vert ^{2} \label{II}%
\end{align}
This expression includes quasiparticle tunneling as well as Andreev processes.
With our non-interacting approach, when $\Gamma_{S}$ increases, subgap Andreev
processes appear much more quickly than what is expected in the Coulomb
blockade regime, because Coulomb interactions forbid $2e$ charge fluctuations
necessary for Andreev reflections\cite{Kang}. In our case, this is not a
problem because we have a low $\Gamma_{S}$. In Fig.\ref{2D}b, the onset of the
non-interacting Andreev current is slightly visible, but this current is
barely above the noise level of the data in the top left panel. For values of
$\Gamma_{S}$ larger than in our experiment, it would be necessary to use an
interacting theory to reproduce satisfactorily the data.

\subsection{Keldysh description of the multisite case\label{multisite}}

One can generalize the approach of Appendices \ref{smcl} and \ref{kel} to
geometries with several quantum dots/sites or several orbitals, denoted with
an index $i$. In the case where each discrete level $i\in\lbrack1,N]$ of the
nanocircuit is shifted by the cavity field $\hat{a}+\hat{a}^{\dag}$ with a
constant $g_{i}$, a semiclassical linear-response description leads to
Eqs.(\ref{tt}), (\ref{Xigg}), and (\ref{CC}) of the main text. These equations
involve generalized advanced and retarded Greens functions $\mathcal{\check
{G}}^{a/r}(\omega)$ which enclose $N\times N$ site/orbital subblocks. The
element $\mathcal{\check{G}}_{ij}^{a/r}(\omega)$ has a Nambu structure:%
\begin{equation}
\mathcal{\check{G}}_{ij}^{a/r}=\left[
\begin{tabular}
[c]{ll}%
$\mathcal{G}_{\hat{d}_{i\uparrow},\hat{d}_{j\uparrow}^{\dag}}^{a/r}$ &
$\mathcal{G}_{\hat{d}_{i\uparrow},\hat{d}_{j\downarrow}}^{a/r}$\\
$\mathcal{G}_{\hat{d}_{i\downarrow}^{\dag},\hat{d}_{j\uparrow}^{\dag}}^{a/r}$
& $\mathcal{G}_{\hat{d}_{i\downarrow}^{\dag},\hat{d}_{j\downarrow}^{c}}^{c}$%
\end{tabular}
\ \ \right]
\end{equation}
with $\mathcal{G}_{A,B}^{a/r}$ scalar Greens functions defined in Appendix B,
and $\check{\Sigma}^{<}$ the lesser self energy of the discrete levels. Above
$\hat{d}_{i\sigma}^{\dag}$ creates an electron with spin $\sigma$ in the
orbital level $i$. The matrix $\check{\tau}_{i}$ is a diagonal matrix which
corresponds to $\check{\tau}$ in the orbital block $(i,i)$ and is zero otherwise.

Note that our formalism assumes that the cavity electric field shifts only the
discrete energy levels $i$. This can be obtained by using AC\ top gates to
reinforce the coupling between cavity photons and the quantum dot. If one uses
a different fabrication technology with e.g. remote AC gates, it can be
necessary to assume that the cavity field also shifts by a different amount
the potentials of the different reservoirs coupled to the dot. In such a case,
one can obtain supplementary effects like for instance a direct influence of
the quantum dot circuit conductance on the cavity linewidth
shift\cite{Delbecq1,Delbecq2,Ares,Dmytruk}. A modulation of tunnel couplings
by the photonic fields could also be relevant for very high tunnel
rates\cite{Cottet2015}. These cases are beyond the scope of the present article.


\begin{thebibliography}{99}
\bibitem {sc}If one uses two N contacts with a similar structure and the same
bias voltage, this only increases the effective density of states $\rho_{N}$
of the normal continuum seen by the quantum dot. In this picture, the
effective tunnel rate $\Gamma_{N}=2\pi\left\vert t_{N}\right\vert ^{2}\rho
_{N}$ corresponds to the sum of the tunnel rates of the two real N contacts.
\bibitem {mapping}This mapping is valid as long as the cavity resonance is
well described by a Lorentzian, which is our case (see Eq.(\ref{tt})). In
particular, the prefactor $t_{0}$, which characterizes our setup transmission,
is assumed to be frequency-independent in the explored range and by
definition, it is not changed by the nanocircuit.
\bibitem {noteS}In the linear limit where $\Delta\varphi$ and $\Delta A$ are
small, one has $\Delta\omega_{0}\simeq\Lambda_{0}\Delta\varphi$ and
$\Delta\Lambda_{0}\simeq-\Lambda_{0}\Delta A/A_{0}$ so that $\Delta\omega_{0}$
and $\Delta\Lambda_{0}$ reveal the cavity frequency and linewidth shifts
respectively. However, since we have strong signals it is more accurate to use
the non linear relations of the main text.
\bibitem {just}In the semiclassical limit, by combining Eqs.(\ref{HH}) and
(\ref{nn}) with $\chi(-\omega_{0})=\chi(\omega_{0})^{\ast}$, one obtains a
photon emission/absorption rate $\Gamma_{e/a}=-2g^{2}\operatorname{Im}%
[\chi(\omega_{0})]n_{ph}$. Then, from Ref.\cite{noteS}, for linear cavity
signals, one has $\Delta\Lambda_{0}=g^{2}\operatorname{Im}[\chi(\omega
_{0})]\simeq-\Lambda_{0}\Delta A/A_{0}$. This gives $\Gamma_{e/a}\simeq
2n_{ph}\Lambda_{0}\Delta A/A_{0}$.
\bibitem {check}As a first test, we have checked that the multisite
Eq.(\ref{Xigg}) agrees with the expressions given in Ref.\cite{Cottet:11} for
a non-interacting equilibrium double quantum dot with N reservoirs.
\bibitem {defXi}In the multidot case we define $\chi_{ij}(t)=-i\theta
(t)\left\langle \{\hat{n}_{i}(t),\hat{n}_{j}(t=0)\}\right\rangle _{0}$ where
$\hat{n}_{i}$ is the quantum operator for the number of electron in site $i$
and $\left\langle {}\right\rangle _{0}$ denotes the statistical averaging with
$g_{i}=0$ for $i\in\lbrack1,N]$
\bibitem {example}For instance, in the case of a non-interacting single
quantum dot in equilibrium with N reservoirs, Eq. (\ref{XiT}) can be recovered
from the Landauer B\"{u}ttiker scattering description used in Ref.\cite{Frey2}.
\end{thebibliography}
\end{document}